\documentclass[aps,prl,twocolumn]{revtex4}
\usepackage{amsbsy,latexsym,amsmath}
\usepackage{amsfonts}
\usepackage{amssymb}
\usepackage[mathscr]{eucal}
\usepackage{epsfig,graphics,graphicx}

\newcommand{\ea}{\emph{et al.}}

\newcommand{\tr}{{\rm tr}\,}

\newcommand{\veta}{{\boldsymbol\pi}}
\newcommand{\vphi}{{\boldsymbol\varphi}}

\begin{document}
\title{Local discrimination of mixed states}

\author{J.~Calsamiglia,  J.~I.~de~Vicente, R.~Mu\~{n}oz-Tapia, E.~Bagan}
\affiliation{Grup d'Informaci\'o Qu\`antica (GIQ), Departament de F\'{\i}sica, Universitat Aut\`{o}noma de
Barcelona, 08193 Bellaterra (Barcelona), Spain}

\begin{abstract}

We provide rigorous, efficiently computable and tight bounds on the average error probability of 
multiple-copy discrimination between qubit mixed states by  Local Operations assisted with Classical 
Communication (LOCC). In contrast to the pure-state case, 
these experimentally feasible protocols perform strictly worse than the general collective ones.
Our numerical results indicate that the gap between LOCC and collective error rates persists in the 
asymptotic limit. In order for LOCC and collective protocols to achieve the same accuracy, the former requires up to twice the number of copies of the latter.
Our techniques can be used to bound the power of LOCC strategies in 
other similar settings, which is still one of the most elusive questions in quantum communication. 
\end{abstract}
\pacs{03.67.Hk, 03.65.Ta}

\maketitle

Quantum communication and computation tasks involve,  broadly speaking, transforming an input state and reading the corresponding output state. One of the most prominent features of quantum mechanics is that, hard as one may try,  
the readout 
will be unavoidably imperfect unless the various \mbox{output}~states are orthogonal. This has both fundamental and practical implications that lie at the heart of quantum mechanics and its applications.
The most simple scenario where this deep fact manifest itself is what is known as  quantum state discrimination or, more generically,  quantum hypothesis testing \cite{bergou04}. In its simplest form,  given one of two possible sources that provide $N$ independent copies of a state~$\rho_{0}$ or $\rho_{1}$ (i.~e., $\sigma_0=\rho_0^{\otimes N}$ or  $\sigma_1=\rho_1^{\otimes N}$), we ask ourselves what is in average the minimum error probability $P_e$ of making a conclusive guess of the identity of the source/state. Hereafter we
will refer to this minimum  average simply as the error probability, and for simplicity we will further assume equal prior probabilities for~$\rho_a$,~$a=0,1$.  State discrimination is an essential primitive for many quantum information tasks, such as quantum cryptography~\cite{crypto}, or even quantum algorithms \cite{bacon05}. Moreover, with the remarkable experimental advances in the preparation and measurement of quantum states it becomes essential to have a theory to assess the performance of state discrimination protocols.

The error probability  for two arbitrary states was given already decades ago 
by Helmstrom~\cite{helstrom}, who provided a formal expression for $P_e$  in terms of the trace distance between the two density matrices~$\sigma_0$ and~$\sigma_1$. An~explicit simple expression can be given for  single qubit states ($N=1$), and only very recently the asymptotic error rate when $N\to\infty$ for qudits has been found through the quantum Chernoff bound~\cite{chernoff1,chernoff2}. For finite (moderately large) number of qubits the permutation invariance of the multiple-copy states~$\{\sigma_a\}$, of size~$\sim2^N$, enables us to write them in block-diagonal form. Thus, we can numerically compute the error probability in terms of the trace-distance between small-sized ($\sim N$) blocks, and the difficulty of the problem (required memory size) becomes  polynomial in~$N$ despite the exponentially growing dimension of the Hilbert space we are dealing with. This collection of results constitute a fairly complete theory regarding the optimal (unconstrained/collective-measurement) multiple-copy discrimination of  quantum states.

The picture changes completely when it comes to discrimination of states using Local Operations (on individual copies) and Classical Communication (LOCC) instead of general (collective) measurements. This scenario is interesting from the fundamental point of view, as it sheds light on the role of quantum correlations in quantum information tasks, but it is of paramount interest from a practical point of view since it puts under scrutiny the attainability of previous bounds in implementations, where collective measurements are usually unfeasible. 
For pure states it has been shown~\cite{brody}  that the minimum collective error probability can be attained by a LOCC one-way adaptive protocol consisting in performing a different von Neumann measurement on each copy, where each measurement is chosen according to the outcome(s) of the previous one(s). Furthermore, it is  shown that an even simpler fixed local strategy, where the \emph{very same} particular von Neumann measurement is repeated on every copy,  though not optimal for finite $N$, does provide the collective asymptotic error rate as $N\to\infty$. The latter also holds when only one of the states is pure  \cite{chernoff2}. 

For mixed states, recently, Hayashi \cite{hayashi09} has proven that, as far as the asymptotic error rate is concerned, one-way adaptive strategies are not advantageous over fixed strategies, which do not make use of classical communication between measurements.  In the last months Higgins \ea~\cite{higgins} have studied theoretically and experimentally the performance of various local strategies including adaptive von Neumann measurements. 
These adaptive strategies, which are optimized using dynamic programming (DP) techniques~\cite{DP}, outperform the others under consideration, but there still remain the fundamental open questions of  whether or not these strategies are the best one can achieve by LOCC (which include generalized local measurements and unlimited communication rounds), and whether or not  those can attain the collective bounds.

In this letter we show how to efficiently compute bounds on LOCC 
protocols. This enables us to compare such protocols to their collective counterparts and benchmark the performance of  particular state discrimination strategies or experiments.  Before presenting our analysis, let us note that there has been some recent interest in the related problem of LOCC discrimination of bipartite states (see \cite{loccEnt} and references therein), where each of the parties has joint access to a share of all copies. We find that in our scenario, although the states are disentangled, quantum correlations play an essential role, which is, in this sense, a manifestation of non-locality without entanglement.

At this point we need to go into a more technical discussion.
The error probability for the case under consideration can be written as~\cite{helstrom}
\begin{equation}
P_e={1\over2}\left\{1+\min_{0\le E\le 1\kern-.22em{\rm l}} \tr[(\sigma_0-\sigma_1) E]\right\}  .
\label{Pe SDP}
\end{equation}
The matrix $E$, together with $\openone-E$, are the elements of the Positive Operator Valued Measure (POVM) that represents mathematically  the measuring protocol. Note that they can be taken to be symmetric under~permutations of the individual systems (invariant~under the action of the symmetric group $S_N$), because so are $\sigma_0$ and $\sigma_1$, and can thus be put in block-diagonal form.

We realize that Eq.~(\ref{Pe SDP}) defines a Semidefinite Programming (SDP) problem, for which very efficient numerical algorithms have been recently developed~\cite{boyd?}.
Bounds on LOCC strategies could be obtained if 
in addition to the positivity constraint $0\le E\le\openone$  on the POVM elements $E$ and $\openone-E$  one would further impose, e.~g., Positive Partial Transposition (PPT), i.~e.,  \mbox{$0\le E^\Gamma\le \openone$} (recall that this condition defines the set of PPT-preserving operations~\cite{Rains}, which includes the set of separable operations and, in turn, LOCC operations). The difficulty here is that Partial Transposition (PT), which we denote by the superscript~$\Gamma$, does {\em not} preserve the block-diagonal form of $E$ (just note that $\Gamma$ breaks permutation invariance).
Hence, the size of the matrices (in particular, $E^\Gamma$) one needs to deal with remains $\sim2^N$ and the error probability cannot be computed but for very small values of~$N$.
We next show how to
go around this problem.

The main observation 
is that
for (two) qubit-state discrimination the POVM element $E$ 
can always be chosen to be PT invariant, $E^\Gamma=E$ (for any bipartite split of the~$N$ qubits). This follows from the fact that with the appropriate choice of basis, one can always cancel all phases in $\rho_0$ and $\rho_1$ simultaneously, so that they have only real elements and are, therefore, symmetric: $\rho_a^{\rm T}=\rho_a$. Obviously, this implies PT invariance, $\sigma_a^\Gamma=\sigma_a$, for any bipartite split.  Hence,  for a given $E$ that satisfies PPT, the PT invariant operator $E'=(E+E^\Gamma)/2$, which also satisfies $0\le E'=E'{}^\Gamma\le\openone$, provides  the exact same error probability~$P_e$. Therefore, we can restrict ourselves to PT invariant operators without any loss of generality.
Since $E$ can be put in block-diagonal form, and there is no need to  apply~$\Gamma$ to check PPT,
the sizes of the matrices we have to deal with grow polynomially  (quadratically) in $N$.

Applying the procedure sketched above  requires, nonetheless,  finding an efficient parametrization of PT invariant matrices in block-diagonal form. The first step towards this end is identifying the independent matrix elements in the computational basis $|i_1i_2\dots i_N\rangle$. For $N$ qubits, the operator $E$ 
can be written as
\begin{equation}
E=\sum_{\{i_p\}}\sum_{\{i'_s\}}E_{i_1i_2\dots i_N}^{i'_1i'_2\dots i'_N} |i_1i_2\dots i_N\rangle\langle i'_1i'_2\dots i'_N| ,
\label{comp basis}
\end{equation}
where all $i_p$, $i'_s$ ($p,s=1,2\dots N$) are either $0$ or $1$ and each sum runs over the $2^N$ possible binary lists (numbers) of~$N$ digits. Invoking permutation invariance and hermiticity
the independent components of $E$ can be chosen to
be~$E_{\kern.1em 11\dots11\dots100\dots00\dots00}^{11\dots 10\dots011\dots10\dots00}\equiv\tilde{\mathscr E}_R^{Q,Q'}$\kern-.3em,
where the first $R$ digits in the subscript are ones and the $Q$  ($Q'$) first digits on top of them (the remaining~$N-R$ zeros) are also ones. We further impose that $R\le Q+Q'$ and note that~$Q\le R$. We next use PT invariance to exchange the last $R-Q$ ones in the subscript with the zeros on top of them by raising (lowering) the corresponding $i_p$ ($ i'_p$). This proves that the PT invariant matrices we are dealing with have~$(N+1)(N+2)/2$ independent components:
\begin{equation}
E_{\kern.1em 11\dots10\dots00\dots00}^{11\dots11\dots10\dots00}\equiv{\mathscr E}_r^{q},\quad
r\le q,
\end{equation}
where $q$ and $r$ are the number of ones in the superscript and the subscript respectively.

We next wish to write $E$ in block-diagonal form. To this end, we map each qubit to a spin~$1/2$, $|i_p\rangle\to|m_p\rangle=|(-1)^{i_p}/2\rangle$, where $m_p$ is the magnetic number of the $p$-th spin, 
and change from the uncoupled (computational) basis to the total spin eigenbasis $\{|j,m\rangle\}_{m=-j}^j$, which span the irreducible representations (irreps) of $SU(2)$.  In this basis $E$ becomes block-diagonal and the matrix elements of each block $E^{(j)}$, i.~e., ${}[E^{(j)}]_m^{m'}=\langle j,m|E|j,m'\rangle$,
are expressed as linear combinations of the independent parameters~${\mathscr E}_r^q$. We write
$\big[E^{(j)}\big]_m^{m'}=\sum_{r,q}\big[{\mathscr M}^{(j)}\big]^{m' r}_{m\,q}{\mathscr E}^q_r$,
which facilitates the SDP implementation of the optimization.
Some comments are in order.
For given~$j$ and~$m$, the state~$|j,m\rangle$ is degenerate. Note, though, that all blocks with the same~$j$ are identical, as $E$ is fully symmetric. Therefore,  the contribution of $E^{(j)}$  to the error probability will have to be multiplied by the corresponding degeneracy,
\raisebox{0em}[.1em][.1em]{$n_j=\big(\!\!\begin{array}{c}\scriptscriptstyle N\\[-.6em] \scriptscriptstyle N/2-j\end{array}\!\!\big)(2j+1)/(N/2+j+1)$}.
The~matrices~${\mathscr M^{(j)}}$ turn out to be (see supplementary material)
\begin{eqnarray}
\big[{\mathscr M}^{(j)}\big]^{m' r}_{m\,q}&=&
\sum_k[\Delta^{(j)}_k]^{m'}_m\begin{pmatrix}
\tfrac{N}{2}-j \\ \tfrac{q-r+m'-m}{2}-k\end{pmatrix}
\nonumber\\[-.3em]
&\times&
(-1)^{\tfrac{q-r+m'-m}{2}-k}\delta_{q+r,N-m-m'} ,
\end{eqnarray}
where we have defined
\begin{equation}
\big[\Delta^{(j)}_k\big]^{m'}_m\!\!=\!{\sqrt{(j\!-\!m)!(j\!+\!m)!(j\!-\!m')!(j\!+\!m')!}\over
(j\!-\!m\!-\!k)!(j\!+\!m'\!-\!k)!(m\!-\!m'\!+\!k)!k!
} .
\label{coeff Wigner}
\end{equation}
In the above expressions the sums run over all integer values for which the factorials make sense.
We note in passing that the very same coefficient~(\ref{coeff Wigner}) appears in the expression of the Wigner d-matrices, which give the different irreps of a rotation about the $y$ axis (see supplementary material).

Now that we have a minimal parameterization of the operators that are invariant under permutations and partial-transpostions we can compute the error probability by the following SDP instance:
\begin{equation}
P_e^{\mathrm{PPT}}\!\!=\!{1\over2}\Big\{\!1\!+\!\min_{\{{\mathscr E}^q_r\}} \sum_{j}n_{j}\tr\left[(\sigma_0^{(j)}-\sigma_1^{(j)}) E^{(j)}\right]\Big\} .
\end{equation}
Here the minimization is constrained
by~\mbox{$0\le E^{(j)} \le \openone$}, for all possible values of the total spin~$j$, and the matrix blocks~$\sigma_a^{(j)}$ are computed to be
\begin{eqnarray}
{}[\sigma_a^{(j)}]_m^{m'}&\!\!\!=\!\!&{\!\left(1\!-\!r_a^2\right)^{\!{N\over2}-j}\over2^N}\!\sum_k\!{}[\Delta_k^{(j)}]^{m'}_m
\!\!\left[\!(-1)^a{r_a\sin\!{\theta\over2}}\right]^{m-m'\!+2k}
\nonumber\\[-.3em]
&\!\!\!\times\!\!&
\bigg(\!{1\!+\!r_a\cos\!{\theta\over2}}\bigg)^{\!j+m'\!-k}
\!\!\bigg(\!{1\!-\!r_a\cos\!{\theta\over2}}\bigg)^{\!j-m-k}\!,
\end{eqnarray}
where $r_a$, often referred to as purity or degree of mixedness, is the length of the Bloch vector $\vec r_a$ of the single qubit state $\rho_a$, and $\theta$ is the relative angle between~$\vec r_0$ and~$\vec r_1$.

As argued above,  $P_e^{\mathrm{PPT}}$ provides a lower bound to the error probability attainable by the most general LOCC strategy, which includes weak generalized local measurements interlaced with an unlimited number of classical communication rounds. In what follows we will compare this bound to the error probability of the optimal collective strategy $P_e^{\rm col}$ and to that of two LOCC strategies that use rank one projective measurements on the individual copies. More precisely:
 i) \emph{Repeated} strategy, where the very same two-outcome measurement is performed on every copy. The error probability, $P_{e}^\mathrm{rep}$, is obtained by minimizing over the azimuthal angle $\Theta$ that specifies the unit Bloch vector of the two measurement projectors. The corresponding asymptotic error rate can also be obtained from the classical Chernoff bound \cite{chernoff2}. 
 ii) \emph{Adaptive}  strategy, where copies are measured sequentially and
the choice of the azimuthal angle $\Theta_s$, corresponding to the measurement on the $s$-th copy,  depends on the outcomes obtained upon measuring the preceding~$s-1$ copies, i.e.,~it makes use of one-way communication. If the number of available copies, $N$, is known beforehand it is possible to find the optimal adaptive strategy  very efficiently using~DP~\cite{higgins}, as detailed in the supplementary material.
\begin{figure}[ht]
\setlength{\unitlength}{.9cm}
\begin{picture}(8,4.7)
\put(-.1,0){  \includegraphics[width=2.6in]{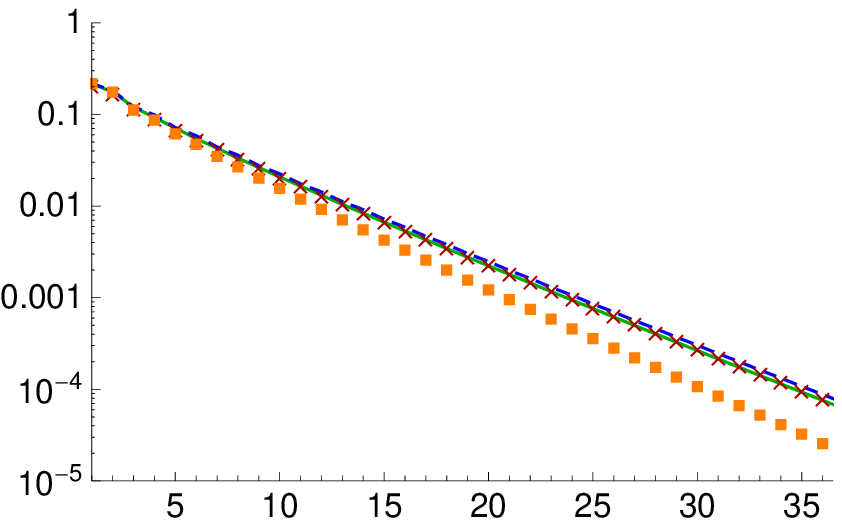} }
\put(3.7,2.4){  \includegraphics[width=3.65cm]{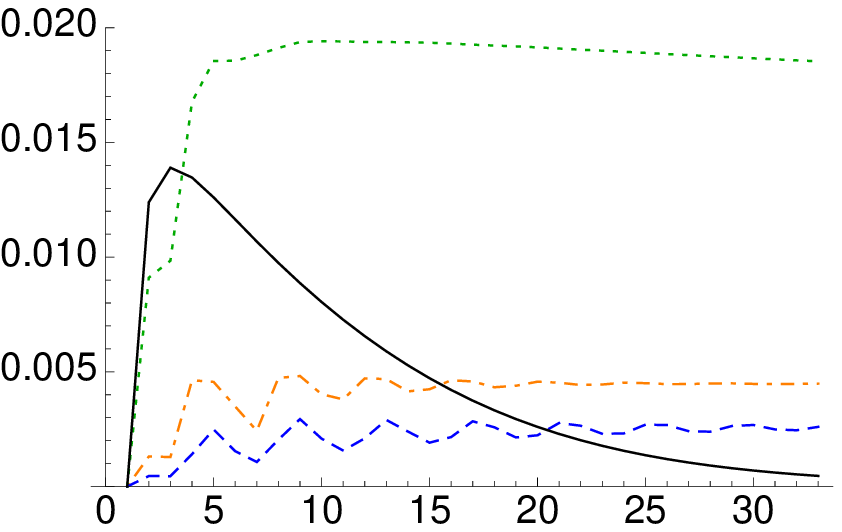} }
\put(.2,4.7){\small $P_e$}
\put(7.5,.2){\small $N$}
\put(6,2.1){\scriptsize $N$}
\put(3.35,3.8){\scriptsize $\Delta$}
\end{picture}
\caption[]{\label{fig:fig1} (color online) LOCC lower-bound $P_e^{\rm PPT}$ (solid line) and error probability for collective (squares), adaptive (crosses) and repeated (dashed) strategies for $r_{0}=r_{1}=.8$ and $\theta=\pi/2$.
Inset: Gap [Eq.~(\ref{gap})] vs.~$N$ for $\theta=\pi/2$,~$r_{0}=.8$ and $r_{1}=1$ (solid), $r_{1}=.9$ (dotted),
$r_{1}=.5$ (dotted-dashed),~$r_{1}=.4$ (dashed).}
\end{figure}

Figure~\ref{fig:fig1} shows the error probability of discrimination between two states with equal purity \mbox{($r_{0}=r_{1}=.8$)} and $\theta=\pi/2$ for the various strategies/bound discussed above. We notice that there is a significant gap between the collective strategy and the LOCC lower-bound~($P_e^{\rm PPT}$). In addition we note that the error probability for repeated and adaptive strategies fall almost on top of the LOCC lower-bound. 
This shows that $P_e^{\rm PPT}$ is a very tight bound and that it can be taken as a good estimate of the minimal LOCC error probability for most practical purposes. The figure clearly shows that, as expected, the error probability falls exponentially with~$N$: $P_{e}\sim \mathrm{e}^{-C N}$.  
By fitting the data in the figure  to an error rate of the form $C=C_{0}+C_{1} \log N/N+C_{2}/N$ for~$25\le N\le 35$, we can obtain its asymptotic value~$C_{0}$ for the various strategies. For the collective and repeated strategies the results agree up to the third significant digit with the analytical results provided by the quantum and classical Chernoff bounds respectively~\cite{chernoff2}.  More interestingly, within numerical accuracy the fits indicate that the gap between collective and LOCC error rates persists in the asymptotic limit.  The results are also consistent with a convergence of the asymptotic error rates for the two LOCC strategies and that of the LOCC lower-bound ($P_e^{\rm PPT}$), although here, due to the already small differences, it is harder to exclude the existence of a (tiny) non-vanishing gap.

In the inset of Figure~\ref{fig:fig1} we plot 
the gap between the collective error rate and that of the LOCC lower-bound~($P_e^{\rm PPT}$), i.~e.: 
\begin{equation}
\Delta=C^{\mathrm{col}}-C^{\mathrm{PPT}}=-{1\over N}\log{P_{e}^{\mathrm{col}}\over P_{e}^{\mathrm{PPT}}}.
\label{gap}
\end{equation}
We notice that 
the gap reaches its asymptotic value already for a small number of copies. This is so for all values of~$r_1$ but for~$r_{1}=1$ (solid line), for which~$\Delta$, after growing to a maximun at~$N\approx4$, decreases to zero  as it should, according to Ref.~\cite{chernoff2}. There, as mentioned at the beginning of this letter, it is shown that when one of the states, say~$\rho_1$, is pure the collective error rate is asymptotically attainable by a repeated strategy. More precisely, by one consisting in performing the measurement defined by 
$E=\rho_1$ on each copy.
The unknown state is claimed to be~$\rho_1$ if the $N$ outcomes of the measurements  correspond to $E$ (none to $\openone-E$), and it is claimed to be~$\rho_0$  otherwise ({\em unanimity vote}).
The asymptotic error rate of this strategy attains the upper-bound $C_{0}\leq-\log{F(\rho_{0},\rho_{1})}$,  where~$F(\rho_{0},\rho_{1})$ is the fidelity, defined as $F(\rho_{0},\rho_{1})=(\tr|\sqrt{\rho_{0}}\sqrt{\rho_{1}}|)^2$ \cite{chernoff2}. For the collective strategy it also holds that $-(1/2) \log F(\rho_{0},\rho_{1})\le C_{0}^\mathrm{col}$.

\begin{figure}[ht]
\setlength{\unitlength}{.95cm}
\begin{picture}(8,4)
\put(.2,-0.1){ \includegraphics[width=6.4cm]{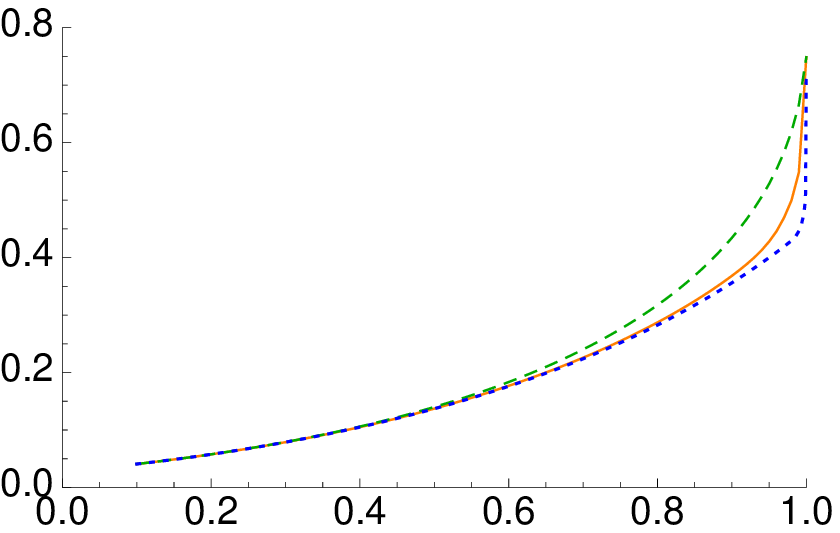} }
\put(1.0,1.6){ \includegraphics[width=3.9cm]{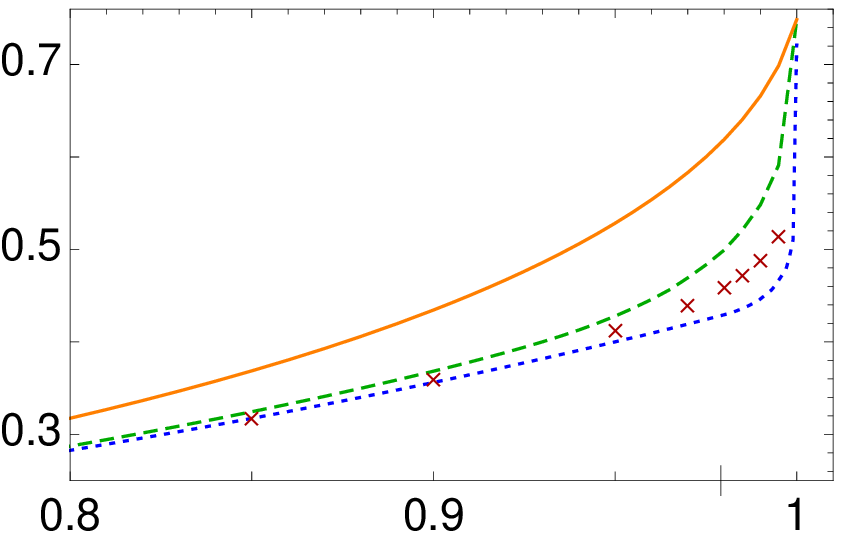} }
\put(-.1,2){$C$}
\put(4.55,1.6){\tiny $r^*$}
\put(4,-.2){$r$}
\end{picture}
\caption[]{\label{fig:fig2}  (b) Error rate vs.~$r=r_0=r_1$ for $N=25$ and $\theta=\pi/2$, and for collective (solid), LOCC (PPT bound) (dashed),  and repeated (dotted) strategies.
Inset shows in addition the error rate for the one-way adaptive strategy~(crosses).}
\end{figure}

Figure~\ref{fig:fig2} shows the error rate $C=-(1/N)\log P_{e}$, for  two equally mixed states, $\theta=\pi/2$ and $N=25$, as one varies their degree of mixedness $r$. We identify four parameter regions in this plot: i) For very mixed states~($r\lesssim .5$) collective and repeated local strategies have essentially the same performance. ii) As the purity increases ($.5\lesssim r\lesssim .8$), the collective strategy starts to outperform the LOCC~one, but the repeated strategy nearly attains the LOCC lower-bound~$P_e^{\rm PPT}$ (upper-bound on the error rate). The measurements of this repeated strategy have~$\Theta=\pi/2$, which means that their Bloch vector is proportional 
to~$\vec r_0-\vec r_1$~\footnote{
This holds for $N$ even, but it is only an approximation for $N$ odd, becoming exact as $N\to\infty$.
}. The decision is taken by ``\emph{majority vote}", i.e.,~the most frequent outcome determines the decision. iii)  At very high purities ($.8\lesssim r\le r^*$) the three LOCC curves start to split.
iv)~At purities larger than a critical one, $r^*$,  all LOCC curves start to rapidly converge to the collective~one. 
At~$r= r^*$ the measurement angle of the repeated strategy starts to change from~$\Theta=\pi/2$ towards~$\Theta=\pi/4$ as~$r\to1$  (for arbitrary~$\theta$, one has~$\Theta\to\theta/2$, i.e.,~the Bloch vector of the measurements goes to either $\vec r_0$ or $\vec r_1$) and the decision rule gradually shifts from a {\em majority vote} to the {\em unanimity vote} described above.   In the asymptotic limit~$N\to\infty$,~$r^*$ can be computed to arbitrary accuracy as a solution of a transcendental equation; for $\theta=\pi/2$ one has $r^*\approx .9819$. For~$r<r^*$ the error rate of the repeated strategy, $C_0^{\rm rep}$,  saturates the fidelity lower-bound introduced above: $C_{0}^\mathrm{rep}=- (1/2) \log F(\rho_{0},\rho_{1})$.  

As mentioned in the introductory part, for asymptotically large~$N$ repeated and one-way adaptive error rates  coincide~\cite{hayashi09}. The tiny gap between the corresponding (dotted) curve and the crosses in Fig.~\ref{fig:fig2} is explained by~the relatively small number of copies ($N=25$) used in the plot.
It is plausible, and consistent with our data, that also the gap between the LOCC error rate bound (dashed line) and $C_{0}^\mathrm{rep}$ vanishes as~$N\to\infty$. 

Note that the smaller the error rate, the more copies we need to achieve the same error probability. The ratio~$f=C^{\mathrm{col}}/C^{\mathrm{PPT}}$ tells us that we need $f N$ copies in order for the best LOCC strategy to discriminate with the accuracy of the collective one. The general features of $f$ as a function of $r$ can be immediately grasped from Fig.~\ref{fig:fig2}.
For very mixed states LOCC and collective strategies require a similar 
number of copies ($f\approx 1$) to discriminate with the same error probability, but as the states become more pure the LOCC strategies demand 
up to twice this number
(i.~e.,~$f\lesssim2$, which asymptotically is the ratio of the two fidelity bounds).
In the limit of very pure states ($r\gtrsim r^*$) the factor $f$ drops back down again to one.

In summary, we have lower-bounded the error probability of LOCC discrimination between two qubit mixed states. To do so, we have characterized all permutation- and PT-invariant operators, which may find application in other quantum communication problems involving~LOCC. Our results indicate an error rate  gap between the best LOCC and collective discrimination protocols that persists as the number of copies goes to infinity.
This gap takes its largest value in the region of nearly pure, but strictly mixed, states. Excluding this region, there are no significant differences in performance between the simplest (repeated) and optimal LOCC strategies. 

We thank E.~Ronco and G.~Via for their contributions in the earlier stages of this work.
We acknowledge financial support from:
the Spanish MICINN, through the Ram\'on y Cajal program~(JC), contract FIS2008-01236, and project QOIT
(CONSOLIDER2006-00019); from the Generalitat de
Catalunya CIRIT, contract  2009SGR-0985; and from Alianza 4 Universidades program (JIdV).

\section*{SUPPLEMENTARY MATERIAL}

Section I of this supplementary material contains details on the derivation of the block-diagonal form of permutation and PT invariant operators on~$({\mathbb C}^2)^{\otimes N}$. 
Section~II provides an example of dynamic programming  applied to state discrimination with one-way LOCC measuremens 

\subsection{I. Permutation and PT invariant operators in block-diagonal form}\label{S1}

Let us consider the permutation invariant operator
\begin{equation}
E=\sum_{\{i_p\}}\sum_{\{i'_s\}}E_{i_1i_2\dots i_N}^{i'_1i'_2\dots i'_N} |i_1i_2\dots i_N\rangle\langle i'_1i'_2\dots i'_N| ,
\label{comp basis}
\end{equation}
where $\{|i_1i_2\dots i_N\rangle\}$ is the (qubit) computational basis 
and each sum runs over the $2^N$ possible sequences of~$N$ bits. 
Permutation invariance implies that $E$ can be written in block-diagonal form (the blocks  corresponding to irreducible tensors) by an appropriate change of basis.
This is 
more familiarly done in the context of spin and/or angular momentum, so we map each qubit to a spin~$1/2$, 
\begin{equation}
|i_p\rangle\to|m_p\rangle=|(-1)^{i_p}/2\rangle, 
\label{map}
\end{equation}
where $m_p$ is the magnetic number of the $p$-th spin. In this, more physical, picture block-diagonalization is achieved by 
the total spin ($j$) eigenbasis $\{|j,m,\alpha\rangle\}_{m=-j}^j$. The index $\alpha$ labels the bases that span the various equivalent irreducible representations (irreps) of~$SU(2)$, each corresponding to a particular way of coupling the individual spins to give the same total spin~$j$. Permutation invariance also implies that the projections of $E$ on these irreducible subspaces, ${\rm span}(\{|j,m,\alpha\rangle\}_{m=-j}^j)$, are identical for all values of $\alpha$ (they depend only on~$j$). Hence, from hereafter, we drop this index and, e.~g., write~${}[E^{(j)}]_m^{m'}=\langle j,m|E|j,m'\rangle$ for the matrix elements of the blocks. 
It follows that,
\begin{eqnarray}
[E^{(j)}]_m^{m'}&\!\!=\!\!&\sum_{\{m_p\}}\sum_{\{m'_s\}}\langle j,m | m_1m_2\cdots m_N\rangle \nonumber\\
&\!\!\times\!\!& \langle m'_1m'_2\cdots m'_N  | j,m'\rangle E_{i_1i_2\cdots i_N }^{i'_1i'_2\cdots i'_N },
\label{change basis}
\end{eqnarray}
where we implicitly use~(\ref{map}) and the sum runs over all possible configurations of the magnetic numbers such that
\begin{equation}
\sum_{p=1}^N m_p=m;\quad \sum_{s=1}^N m'_s=m'. 
\label{mag num cons}
\end{equation}

Since all the matrix blocks with the same $j$ are identical, to compute the values of the (generalized) Clebsch-Gordan coefficients $\langle m_1m_2\dots m_N|j,m\rangle$ in~(\ref{change basis}) 
it suffices to couple the individual spins in the way that leads to the simplest calculation. For a given $j$, it proved useful to pair in singlets the $N-2j$ spins/qubits that are at the beginning of the sequence and couple the remaining~$2j$ to give total spin~$j$. 
%
%
Thus, we take the state $|j,m\rangle$ to be of the form
\begin{equation}
|j,m\rangle=| \psi_{-}\rangle^{\otimes(N/2-j)}|j,m\rangle_{\mathrm{sym}}  ,
\end{equation}
where $|\psi_-\rangle$ is the spin singlet [$|\psi_-\rangle=(|01\rangle-|10\rangle)/\sqrt2$ in binary notation] and  $|j,m\rangle_{\mathrm{sym}}$ is the fully symmetric state of $2j$ spins with total spin $j$.  With this choice, the only non-vanishing Clebsch-Gordan coefficients are those for which the first $N-2j$ spins in $\langle m_1m_2\dots m_N|$ consist of~$l$ pairs of the form $\langle\frac{1}{2}-\kern-.3em\frac{1}{2}|\equiv\langle\sigma|$  and $p$ pairs of the form~$\langle-\kern-.1em\frac{1}{2}\;\frac{1}{2}|\equiv\langle\mu|$
($01\equiv\sigma$ and $10\equiv\mu$ in binary notation) with $l+p=N/2-j$. 
In this case, and further assuming that conditions~(\ref{mag num cons}) hold, a straightforward calculation yields
\begin{equation}
\langle m_1m_2\dots m_N | j,m\rangle\!=\!(-1)^p\!\left(\!{1\over\sqrt2}\!\right)^{\!\!\tfrac{N}{2}-j}
\!\!\!\left(\!
                                                                                         \begin{array}{c}
                                                                                           2j \\
                                                                                           j+m \\
                                                                                         \end{array}
                                                                                         \!\!
                                                                                       \right)^{\!-\tfrac{1}{2}}\kern-.5em,
\label{CGC}
\end{equation}
where we have used that the fully symmetric state~$|j,m\rangle_{\mathrm{sym}}$ is an equal superposition of 
\raisebox{0em}[0em][.4em]{$\big(\!\!\begin{array}{c}\scriptscriptstyle 2j\\[-.6em] \scriptscriptstyle j+m\end{array}\!\!\big)$} vectors of the form $\otimes_{p=N-2j}^N|m_p\rangle$.

We can use again permutation invariance to cast the independent components of $E$ into the form
\begin{equation}
E_{\,\underbrace{\scriptstyle\sigma\ldots\sigma\;\sigma\ldots\sigma}_{\scriptstyle l+p}\kern.05em\underbrace{\scriptstyle11\ldots1\kern.15em1\ldots1}_{\scriptstyle R}00\ldots0\kern.1em0\ldots00}^{\overbrace{\scriptstyle\sigma\ldots\sigma}^{\scriptstyle l}
\overbrace{\scriptstyle\mu\ldots\mu}^{\scriptstyle p}\overbrace{\scriptstyle11\ldots1}^{\scriptstyle Q}0\ldots0\overbrace{\scriptstyle11\ldots1}^{\scriptstyle Q'}0\ldots00}\equiv\widetilde{\mathcal{E}}_R^{\;l,p;Q,Q'},
\label{def calE}
\end{equation}
where $l$, $p$, $R$, $Q$ and~$Q'$ stand for the number of \mbox{$\sigma$-pairs}, $\mu$-pairs and ones indicated by the braces.
For a fixed choice of~$l$ and~$p$, the unpaired bits are arranged so that in the subscript sequence (the row index of $E$) no zero precedes a one. 
The first $R$ bits on the right of the $\mu$-pairs in the superscript sequence  
(column index) are arranged along the same pattern, and so are the bits placed on top of the $N-l-p-R$ zeroes of the subscript sequence. Note that~$Q\le R$ and
%
%
%
\begin{equation}\label{forms}
\widetilde{\mathcal{E}}_R^{\;l,p;Q,Q'}=\widetilde{\mathcal{E}}_{R+l+p}^{\;0,0;Q+l,Q'+p}\equiv\widetilde{\mathcal{E}}_{R+l+p}^{\;Q+l,Q'+p}.
\end{equation}
It is a straightforward exercise in combinatorics to compute the number of matrix elements of $E$ whose value is~$\widetilde{\mathcal{E}}_R^{\;l,p;Q,Q'}$. It is given by
\begin{eqnarray}
2^{l+p}\!
\left(\begin{matrix} l\!+\!p\\ l\end{matrix}\right)\!\!
\left(\begin{matrix} N\!-\!2l\!-\!2p\\ R\!\end{matrix}\right)\!\!
\left(\begin{matrix} R\\ Q\end{matrix}\right)\!\!
\left(\begin{matrix} N\!-\!2l\!-\!2p\!-\!R\\ Q'\end{matrix}\right) .
\label{combinatorics}
\end{eqnarray}
%
%
The first factor takes into account that by applying a permutation (thus leaving $E$ unaltered) we can replace any of the $l+p$ $\sigma$-pairs in the subscript by a $\mu$-pair (with an exchange $\mu\leftrightarrow\sigma$ at the exact same position of the superscript sequence). The second factor is the number of ways the~$l$ $\sigma$-pairs (and the~$p$ $\mu$-pairs) can be placed  in the first $l+p$ positions of the superscript sequence. The third factor gives the different ways the $R$ ones in the subscript sequence can be placed in $N-2l-2p$ positions  (with a corresponding repositioning of the bits right on top of them). Similarly, the forth (fifth) factor is the number of ways $Q$ ($Q'$) ones can be placed on top of the~$R$ ones ($N-2l-2p-R$ zeroes) of the subscript sequence. 
                                
With the Clebsch-Gordan coefficients~(\ref{CGC}), the notation introduced in~(\ref{def calE}),  and the combinatorics in~(\ref{combinatorics}), we can give an explicit expression for the matrix-blocks $E^{(j)}$ as 
%
%
\begin{eqnarray}
&&\kern-2em{}[E^{(j)}]_m^{m'}=
{2^{j-\tfrac{N}{2}}}\sqrt{
\left(\!
                                                                                         \begin{array}{c}
                                                                                           2j \\
                                                                                           j+m \\
                                                                                         \end{array}
                                                                                         \!\!
                                                                                       \right)
                                                                                       \left(\!
                                                                                         \begin{array}{c}
                                                                                           2j \\
                                                                                           j+m' \\
                                                                                         \end{array}
                                                                                         \!\!
                                                                                       \right)
}\nonumber\\
&&\kern-2em\times\sum_{l=0}^{\tfrac{N}{2}-j}\;\;\sum_{k=0} ^{j-m}(-1)^{\tfrac{N}{2}-j-l}\; \widetilde{\mathcal{E}}_{j-m}^{\;l,\tfrac{N}{2}-j-l; \,k,j-m'-k}
\nonumber\\[.5em]
&&\kern-2em\times\;\,
2^{\tfrac{N}{2}-j}\!\!
\left(\begin{matrix}\! \tfrac{N}{2}\!-\!j\\ l\end{matrix}\right)\!\!
\left(\begin{matrix} 2j\\ j\!-\!m\!\end{matrix}\right)\!\!
\left(\begin{matrix} j\!-\!m\\ k\end{matrix}\right)\!\!
\left(\begin{matrix} j+m\\ j\!-\!m'\!\!-\!k\end{matrix}\right),
 \label{blocks}
\end{eqnarray}
%
%
where we have used that $p=N/2-j-l$, $R=j-m$, $Q'=j-m'-Q$, and finally replaced the dummy index $Q$ by $k$.
%
Using Eq.~(\ref{forms}) and making the changes $k\to j-m-k$ and $l\to N/2-j-l$, this last expression reduces to
%
\begin{eqnarray}\label{Ej-Perm}
{}[E^{(j)}]_m^{m'}&=&\sum_{k}[\Delta_k^{(j)}]_m^{m'}
\sum_{l}
\left(\!\!
  \begin{array}{c}
    \frac{N}{2}-j \\
    l \\
  \end{array}
  \!
\right)\nonumber\\
&\times&(-1)^l\;
\widetilde{\mathcal{E}}_{\tfrac{N}{2}-m}^{\;\tfrac{N}{2}-m-k-l,m-m'+k+l},
\end{eqnarray}
where we have defined
\begin{equation}
\big[\Delta^{(j)}_k\big]^{m'}_m\!\!=\!{\sqrt{(j\!-\!m)!(j\!+\!m)!(j\!-\!m')!(j\!+\!m')!}\over
(j\!-\!m\!-\!k)!(j\!+\!m'\!-\!k)!(m\!-\!m'\!+\!k)!k!
} ,
\label{coeff Wigner}
\end{equation}
and the sums run over all integer values for which the factorials make sense.

Partial transposition invariance  implies that
\begin{equation}
 \widetilde{\mathcal{E}}_R^{\;Q,Q'}=\widetilde{\mathcal{E}}_Q^{\;0,Q'+R}\equiv\mathcal{E}_Q^{Q'+R},
 \end{equation} 
and we can further simplify the expression of the matrix blocks $E^{(j)}$,
%
\begin{eqnarray}\label{Ej-PTinv}
{}[E^{(j)}]_m^{m'}&=&\sum_{k}[\Delta_k^{(j)}]_m^{m'}
\sum_{l}
\left(\!\!
  \begin{array}{c}
    \frac{N}{2}-j \\
    l \\
  \end{array}
  \!
\right)\nonumber\\
&\times&(-1)^l\;
\mathcal{E}_{\tfrac{N}{2}-m-k-l}^{\tfrac{N}{2}-m'+k+l},
\end{eqnarray}
from which Eq.~(4) in the main text follows immediately.

Eqs.~(\ref{Ej-Perm}) 
provides a useful parameterization of general permutation invariant operators. As an example, we compute the block decomposition of the product state $\sigma=\rho^{\otimes N}$, which is also used in the main text [Eq.~(7)].
The independent components of $\sigma$ are easily found using~(\ref{def calE}) [here we do not use superscripts; we write $\rho_{ij}$ ($i,j=0,1$) instead of $\rho_i^j$ for the matrix elements of $\rho$]
\begin{equation}
\widetilde{\mathcal{E}}_R^{\;Q,Q'}=\rho_{11}^Q\rho_{10}^{R-Q}\rho_{01}^{Q'}\rho_{00}^{N-R-Q'} .
\label{param rho}
\end{equation}
Substituting in Eq.~\eqref{Ej-Perm} we obtain
%
\begin{eqnarray}\label{sigma^j}
{}[\sigma^{(j)}]_m^{m'}&=&\sum_{k}[\Delta_k^{(j)}]_m^{m'}
\sum_{l}
\left(\!\!
  \begin{array}{c}
    \frac{N}{2}-j \\
    l \\
  \end{array}
  \!
\right)\nonumber\\
&\times&(-1)^l\;
\rho_{10}^k \rho_{01}^{m-m'+k} \rho_{11}^{j-m-k} \rho_{00}^{j+m'-k}\nonumber\\[.4em]
 &\times&( \rho_{10} \rho_{01})^{l}( \rho_{11} \rho_{00})^{\tfrac{N}{2}-j-l} .
\end{eqnarray}
%
%
Note that the sum over $l$  can be readily performed to give $( \rho_{00} \rho_{11}- \rho_{10} \rho_{01})^{N/2-j}=(\det \rho)^{N/2-j}$, and
%
\begin{eqnarray}\label{sigma^j sim}
{}[\sigma^{(j)}]_m^{m'}&=&(\det \rho)^{N/2-j}\sum_{k}[\Delta_k^{(j)}]_m^{m'}
\nonumber\\
&\times&
\;
\rho_{10}^k \rho_{01}^{m-m'+k} \rho_{11}^{j-m-k} \rho_{00}^{j+m'-k}
\end{eqnarray}
Eq.~(7) in the main text follows from the Bloch parameterization of a rebit used there
\begin{eqnarray}
&\displaystyle \rho_{00}={1+r\cos(\theta/2)\over2},\quad\rho_{11}={1-r\cos(\theta/2)\over2},&
\nonumber\\
&\displaystyle  \rho_{01}=\rho_{10}={ r\sin(\theta/2)\over2}.&
\end{eqnarray}
%

Finally, Eq.~(\ref{Ej-Perm}) can also be used to provide an alternative derivation of the well-known  Wigner d-matrices. Consider the block decomposition of the product representation $d^{\otimes N}$ of the $SU(2)$ ``rotation" about the $y$-axis: 
\begin{equation}
d_{00}\!=\! \cos\frac{\beta}{2},\ d_{11}\!=\! \cos\frac{\beta}{2},\ d_{01}\!=\!-d_{10}\!=\!-\sin\frac{\beta}{2}.
\end{equation}
%
Proceeding as above [Eqs.~(\ref{param rho}) to~(\ref{sigma^j sim})] one obtains
%
\begin{align}
[d^{(j)}(\beta)]_m^{m'}\!&=\!\sum_k[\Delta_k^{(j)}]_m^{m'}\left(-1\right)^{m-m'+k}\nonumber\\
&\!\times\!
\left(\!\cos\tfrac{\beta}{2}\!\right)^{2j-m+m'\!-2k}
\!\left(\!\sin\tfrac{\beta}{2}\!\right)^{m-m'\!+2k},
\end{align}
which are the standard Wigner d-matrices.

\subsection{II. One-way LOCC State discrimination using dynamic programming}\label{S2}

Dynamic Programming (DP) is 
a powerful numerical technique with a wide range of applicability that, roughly speaking, consists in casting an optimization problem into a recursive form. It can be applied to problems~where decisions that involve smaller-sized \mbox{sub-problems} are taken at different stages of the optimization process.  Its main advantage is that it avoids exploring those decisions that cannot possibly be optimal.  Here, we present a DP formulation of multiple-copy discrimination among $M$ possible states~$\{\rho_k\}_{k=1}^M$ using a local one-way adaptive strategy. It is a slight generalization of the problem considered in the main text, for which~$M=2$.

Let us assume that we are given $N$ identical copies of a state that is known to belong to the set~$\{\rho_k\}_{k=1}^M$, and let~$\pi(k)$ be the prior probability (prior for brevity) that the given state be~$\rho_k$; for convenience we write~$\veta=\{\pi(k)\}_{k=1}^M$. We view $\veta$ as a free variable, which can take arbitrary values in ${}[0,1]^M$. The discrimination protocol we wish to optimize consists in measuring each copy sequentially, and returning the alleged identity of the given state after completion of the whole sequence, once the \mbox{$N$-th} copy has been measured.
When an outcome is obtained, we consequently update the priors and perform a new measurement on the next copy. This new measurement is specifically chosen in accordance with the just updated priors. 

For simplicity, let us here assume that the POVM operators  that characterize the local measurements have the form~$\{E_\chi(\vphi)\}_{\chi=1}^W$, i.~e., they can be parameterized by a set of variables 
which we have arranged into the vector~$\vphi$, and $W$ is the number of possible outcomes. 
%
%
If the unknown state is $\rho_k$ the (conditioned) probability of obtaining the outcome~$\chi$ when performing the measurement on a copy is given by the Born rule 
\begin{equation}
p(\chi|k;\vphi)=\tr[E_\chi(\vphi)\,\rho_k],
\label{Born rule}
\end{equation} 
where a semicolon will always be used to separate random variables from any other argument the various probabilities may depend upon.
As mentioned above, on each copy we allow for a specific measurement depending on the current value of the priors.
Our aim is to optimize the function $\vphi(\veta)$ at each step of the measurement sequence so that the success probability of discrimination after the $N$-th (last) measurement is as large as possible. We will denote the optimal choices by $\vphi_n^*(\veta)$, where $n=1,2,\dots,N$ refers to the order in the sequence and the asterisk ($*$) stands for ``optimal".  In order to do so in a recursive way, we still need to introduce  a further probability function, which we denote by $S^*_{n-1}(\veta)$. It is defined to be the maximum average success probability if we drop the first $n-1$ copies, namely, if the sequence of measurements starts at copy $n$, and assuming that the priors are~$\veta$.
Again, the asterisk reminds us that this sequence of~$N-n+1$ measurements is optimal for the given priors. Notice that all information about the identity of the unknown state that we would gain by performing measurements on the first $n-1$ copies is encoded in the priors $\veta$ (i.~e., in the posterior probabilities if the measurements on the first $n-1$ copies were performed). Notice also that the success probability we wish to compute is given by $S_0^*(\veta)$, and that
\begin{equation}
S^*_N(\veta)=\max_k\{\pi(k)\}.
\label{initial cond}
\end{equation}
In words, this equation states that after performing the last measurement, the optimal decision is to identify the unknown state as being the state in~$\{\rho_k\}_{k=1}^M$ with maximum posterior probability (Bayes decision). 

We next establish a recursion relation that  will give us~$S^*_{n-1}(\veta)$ once we know $S^*_n(\veta)$. 
 %
%
%
After performing the measurement on the $n$-th copy, and assuming we have obtained outcome~$\chi$, the well known Bayes rule enables us to update the priors as $\veta\to\veta'(\chi;\veta,\vphi)$, with $k$-components given by:
\begin{equation}
\pi'(k|\chi;\veta,\vphi)=\frac{p(\chi|k;\vphi)\pi(k)}{p(\chi;\veta,\vphi)}.
\label{Bayesian update}
\end{equation}
Here the marginal probability in the denominator (the probability of obtaining outcome~$\chi$) is 
\begin{equation}
p(\chi;\veta,\vphi)=\sum_k p(\chi|k;\vphi)\pi(k) ,
\label{Bayes rule}
\end{equation}
and we explicitly show all the dependencies involved in the updating. The success probability for this sequence of  measurement starting with the $n$-th copy satisfies
\begin{equation}\label{recurr}
S_{n-1}(\veta,\vphi)=\sum_{\chi}p(\chi;\veta,\vphi)\, S_n^*\big(\veta'(\chi;\veta,\vphi)\big);
\end{equation}
each term in this sum is the (joint) probability of obtaining outcome~$\chi$ in the~$n$-th measurement and succeeding at the end of the discrimination protocol. Hence, this sum gives the success probability $S_{n-1}(\veta,\vphi)$. Note, however, that it still depends on our measurement choice through $\vphi$ and optimality demands that we maximize this expression over all possible values of these parameters, i.~e.,
\begin{equation} 
S^*_{n-\!1}\!(\veta)\!=\!\max_{\vphi} S_{n-\!1}\!(\veta,\vphi)\!=\!S_{n-\!1}\!\big(\veta,\vphi^*_n(\veta)\big),
\label{S -> S*}
\end{equation}
where the last equation defines $\varphi_n^*(\veta)$ as the choice of parameters that maximize $S_{n-1}(\veta,\vphi)$.
%
%
Eqs.~(\ref{Bayesian update}) to~(\ref{S -> S*}) enable us to recursively compute the success probability~$S_0^*(\veta)$ starting with the initial condition~(\ref{initial cond}). This method is efficient as only the values of $S^*_n(\veta)$ are needed to obtain $S^*_{n-1}(\veta)$ and these can be stored at each step of the optimization.

We should point out here that in general the optimal success probability $S^*_n(\veta)$ and measurement parameter functions $\vphi^*_n(\veta)$ can only be computed numerically, which requires discretizing $[0,1]^{M}$ with a fairly large set of lattice points and using interpolation methods. For two qubits discrimination with von Neumann measurements~$\vphi$ is one-dimentional, i.~e., it is only the azimuthal angle on the Bloch sphere, and so is $\veta$, as $\pi(1)=1-\pi(0)$, and~DP is very efficient. In this simple case and with equal priors the error probability is 
\begin{equation}
P_e=1-S_0^*(1/2). 
\end{equation}
We have used a one dimensional lattice of~20,000 points to compute our results and  checked that they are robust against changes of the number of lattice points. 

Having explained the DP technique, one may wonder whether or not this is indeed the most general one-way adaptive local strategy. To show that this is so, we formally iterate~Eqs.~(\ref{Bayesian update}) to~(\ref{S -> S*}) and obtain
%
%
\begin{eqnarray}
S^*_0(\veta)&=&\max_{\vphi_{1}}\sum_{\chi_{1}}\max_{\vphi_{2}}\sum_{\chi_{2}}\cdots\max_{\vphi_{N-1}}\sum_{\chi_{N-1}}\max_{\vphi_{N}}\sum_{\chi_{N}}\nonumber \\
&\times& \max_{k}\left\{\pi(k)\prod_{n=1}^N p(\chi_{n}|k;\vphi_{n})\right\} .
\label{wonder}
\end{eqnarray}
The argument of $\max_k$ in~(\ref{wonder}) is~$p(\{\chi_n\},k;\veta,\{\vphi_n\})$, the joint probability of obtaining a sequence of outcomes~$\{\chi_n\}_{n=1}^N$ ($\{\chi_n\}$ in short, and similarly for~$\{\vphi_n\}$), and the unknown state being~$\rho_k$. The maximization over~$k$ corresponds to the Bayes decision: given an outcome, or sequence of outcomes, decide for the state~$\rho_k$ that gives that outcome with highest probability.
The sum over all outcomes gives the total (or average) success probability. 
Note that the maximization over measurement parameters is performed in a specific order. Because of this,
the choice of~$\vphi_1$ is independent of the outcomes $\{\chi_n\}$. However, since the sum over~$\chi_1$ precedes~$\max_{\vphi_2}$, we can make the second measurement
depend on the outcome of the first one, i.e. $\vphi_2(\chi_1)$. In general, $\vphi_n(\{\chi_m\}_{m=1}^{n-1})$ (the $n$-th measurement depends on the outcomes obtained in the previous ones). Hence we find that, indeed, the DP procedure returns the optimal one-way LOCC strategy (within the family of local POVM's parametrized by $\vphi$).

Note that if we were to follow a brute force method and make direct use of~(\ref{wonder}), we would be bound to optimize over an impractically large number of parameters~$\sim \exp N$, instead of the~$\sim N$ optimization parameters present in the equivalent DP formulation.

\end{document}